\documentclass[11pt]{article}

\usepackage{amsmath,amssymb}

\usepackage{epsfig}
\usepackage{graphicx}               
\usepackage{url}
\usepackage{hyperref}

\newcommand{\nc}{\newcommand}

\nc{\beq}{\begin{equation}}
\nc{\eeq}{\end{equation}}
\nc{\beqa}{\begin{eqnarray}}
\nc{\eeqa}{\end{eqnarray}}
\nc{\bea}{\begin{eqnarray}}
\nc{\eea}{\end{eqnarray}}
\nc{\ra}{\rightarrow}
\nc{\lsim}{\begin{array}{c}\,\sim\vspace{-21pt}\\< \end{array}}
\nc{\gsim}{\begin{array}{c}\sim\vspace{-21pt}\\> \end{array}}
\nc{\Tr}{{\rm Tr}}
\nc{\slsh}{\slash\hspace*{-0.22cm}}

\def\be{\begin{equation}}
\def\ee{\end{equation}}
\def\bea{\begin{eqnarray}}
\def\eea{\end{eqnarray}}
\def\bit{\begin{itemize}}
\def\eit{\end{itemize}}

\newcommand{\met}{E^{\rm miss}_T}
\newcommand{\vecmet}{\vec{E}^{\rm miss}_T}

\def\to{\rightarrow}

\title{
\vspace*{-2.3cm}
\begin{flushright}
\normalsize{
SLAC-PUB-14897
  }
\end{flushright}
\vspace{1.5cm}
\Large
\textbf{
Stop the Top Background of the Stop Search
 \\
}\vspace*{1.0cm}
}

\author{Yang Bai$^{a}$, Hsin-Chia Cheng$^{b}$, Jason Gallicchio$^{b}$, and Jiayin Gu$^{b}$
\vspace{5mm}
\\
$^{a}$ \normalsize\emph{SLAC National Accelerator Laboratory, 2575 Sand Hill Road, Menlo Park, CA 94025, USA} \\
$^{b}$ \normalsize\emph{Department of Physics, University of California, Davis, CA 95616, USA}
}

\date{}

\begin{document}
\setcounter{page}{0}
\maketitle

\vspace*{1cm}
\begin{abstract}
The main background for the supersymmetric stop direct production search comes from Standard Model $t \bar t$ events. For the single-lepton search channel, we introduce a few kinematic variables to further suppress this background by focusing on its dileptonic and semileptonic topologies. All are defined to have end points in the background, but not signal distributions. They can substantially improve the stop signal significance and mass reach when combined with traditional kinematic variables such as the total missing transverse energy. Among them, our variable $M^W_{T2}$ has the best overall performance because it uses all available kinematic information, including the on-shell mass of both $W$'s. We see 20\%--30\% improvement on the discovery significance and estimate that the 8 TeV LHC run with 20 fb$^{-1}$ of data would be able to reach an exclusion limit of 650--700 GeV for direct stop production, as long as the stop decays dominantly to the top quark and a light stable neutralino. Most of the mass range required for the supersymmetric solution of the naturalness problem in the standard scenario can be covered.
\end{abstract}

\thispagestyle{empty}
\newpage

\setcounter{page}{1}

\baselineskip18pt

\vspace{-3cm}
\section{Introduction}
\label{sec:introdction}
A main goal of the Large Hadron Collider (LHC) experiments is to understand electroweak symmetry breaking. In the Standard Model (SM), it is achieved by the vacuum expectation value (VEV) of a scalar Higgs field. However, a fundamental scalar field receives quadratically divergent radiative contribution to its mass-squared and suffers from the hierarchy problem. One of the most promising solutions to the hierarchy problem is supersymmetry (SUSY) which introduces a superpartner to every SM field, so that the quadratically divergent corrections to the Higgs mass-squared can be canceled between SM particles and their superpartners. Supersymmetry has been extensively searched for at colliders, and so far we have not found any evidence for it. The latest LHC search results constrain the masses of the gluino and (light generation) squarks in the minimal supergravity~\cite{Chamseddine:1982jx,Barbieri:1982eh,Hall:1983iz} or constrained minimal supersymmetric standard model (MSSM)~\cite{Kane:1993td} to be greater than about 1\,TeV~\cite{Chatrchyan:2011zy,Aad:2011ib}. At face value, it may imply a serious fine-tuning of the electroweak scale if SUSY exists. However, as the largest radiative correction to the Higgs mass in the SM comes from the top quark loop, only the top superpartners (stops) need to be light enough to cancel the top loop contribution~\cite{Dimopoulos:1995mi, Cohen:1996vb}. The gluino and first two generation squarks can be heavier than 1\,TeV without a naturalness problem, at least at one-loop level~\cite{Dimopoulos:1995mi, Cohen:1996vb, Kitano:2006gv, Barbieri:2009ev, Brust:2011tb, Papucci:2011wy}. Therefore, searching for the top superpartners at the LHC offers the most important test of whether SUSY provides a natural solution to the hierarchy problem.

Many third generation squark searches at the LHC rely on gluino production, with subsequent decay to stops or sbottoms~\cite{Aad:2011ks,Chatrchyan:2011bj, Kane:2011zd, Essig:2011qg, Berger:2011af}. This is because the production cross section for gluino is much larger than for direct stop or sbottom production, as long as the gluino mass is not much heavier. However, since the naturalness of the electroweak breaking scale does not require gluino to be light enough to be copiously produced at the 7 or 8 TeV LHC, a more robust stop search would only rely on direct stop pair production. In this paper we focus on the stop search in this channel in a standard $R$-parity conserving SUSY scenario. Here the stop decays to a top quark and the lightest supersymmetric particle (LSP), which is assumed to be a neutralino. Although a light neutralino is not required by naturalness, it avoids the stable charged particle problem and provides a natural candidate for dark matter. The signal we are looking for is $t\bar{t} + \met$ where the missing transverse energy $\met$ comes from the pair of neutralino LSP's which escape the detector. We also assume that the mass difference between the stop and the LSP is substantially larger than the top quark mass. Otherwise the signal will strongly overlap with the SM backgrounds, which would need some different search strategies~\cite{Carena:2008mj,Bornhauser:2010mw,Kats:2011it,He:2011tp,Drees:2012dd}.

In fact, the $t\bar{t}+\met$ signal also occurs in many other extensions of the SM that include a dark matter candidate. It is quite natural to mitigate the hierarchy problem with a relatively light top ``partner'' that decays to the top quark and the dark matter particle. This is possible if the partner is also charged under the symmetry that protects the stability of the dark matter particle. Examples are the little Higgs models with $T$-parity~\cite{Cheng:2003ju,Cheng:2004yc,Cheng:2005as, Bai:2008cf}, models with the exotic fourth generation and dark matter~\cite{Alwall:2010jc}, models with gauged baryon and lepton numbers~\cite{FileviezPerez:2010gw}, and so on. Consequently, the same search applies to many different models, but the mass-reach depends on each model's top partner production cross section. Studies of the $t\bar{t} + \met$ signal for new physics have been performed by many groups in various (fully hadronic, single-lepton) channels in recent years~\cite{Cheng:2005as,Alwall:2010jc,Meade:2006dw,Matsumoto:2006ws,Kong:2007uu,Han:2008gy,Matsumoto:2008fq,Plehn:2010st,Plehn:2011tf}.

The ATLAS collaboration at the LHC has done such a search in the single-lepton channel based on 1.04\,fb$^{-1}$ of data~\cite{Aad:2011wc}. In the single-lepton channel, one $W$ from the top decays leptonically and the other $W$ decays hadronically. Requiring one lepton in the final state suppresses QCD multijet backgrounds tremendously while still retaining a significant $W$ branching fraction. The final state signal consists of four (or more) jets (including two $b$-jets), one lepton, and missing transverse energy. Besides the standard transverse momentum $p_T$ and pseudo-rapidity $|\eta|$ requirements for each object, the two main variables used for separating the signal and backgrounds are the missing transverse energy $\met$ and the transverse mass $M_T$ constructed from the lepton and $\met$~\cite{Aad:2011wc}. The signal events are expected to have large $\met$ from the top-partner decays, along with a neutrino from the leptonic $W$. A hard cut on $\met$ very effectively reduces the SM backgrounds. The cut on $M_T$ removes the backgrounds where the $\met$ is mostly due to a single neutrino from a $W$ decay because the $M_T$ distribution has an end point at $M_W$ in such cases. The existing ATLAS study focused on a fermionic top partner and can exclude this partner's mass up to 420\,GeV~\cite{Aad:2011wc}. There was no sensitivity to the SUSY stop with this limited amount of data because of the much smaller cross section for the scalar particles. Last year's run already delivered more than 5\,fb$^{-1}$ of data. This year, the LHC is expected to deliver even more luminosity at a higher center of mass energy of 8\,TeV. Given the importance of the stop (and other top-partner) search, it is desirable to extend the mass reach using current and future data.

The single-lepton channel analysis of the ATLAS top partner search paper~\cite{Aad:2011wc} found that the largest background remaining after the their cuts on $\met$ and $M_T$ is the dileptonic $t\bar{t}$. In these background events, both $W$'s decay leptonically, but one of the leptons is not reconstructed, is outside the detector acceptance, or is a $\tau$ lepton (which may be misidentified as a jet). Each event contains at least two neutrinos that can produce a large $\met$ and also make it easier to pass the $M_T$ cut. The additional jets come from QCD initial state radiation (ISR). The next-to-largest background comes from the semileptonic $t\bar{t}$ and $W$+jets. The other backgrounds are small after the $\met$ and $M_T$ cuts. To improve the search reach, we designed kinematic variables to identify $t\bar{t}$ backgrounds, focusing on its decay topology. We find that the signal significance and mass reach can indeed be substantially improved with the help of these variables. This paper is organized as follows. In the next section we discuss several such variables. In section~\ref{sec:performance}, we compare the performances of the basic set of cuts and cuts including the new variables, identifying an economical set.

\section{Kinematic Variables for the $t\bar{t}$ Backgrounds}
\label{sec:MT2}

We study the LHC search for the pair-production of stops, $p p \rightarrow \tilde{t}_1 \tilde{t}_1^*$, with $\tilde{t}^0_1 \rightarrow t + \tilde{\chi}_1$.\footnote{We focus on the light mass eigenstates, and calculate reach based on 100\% decay to $t+\tilde{\chi}^0$.}  We focus on the signal's one-lepton decay channel, in which one top quark decays leptonically $t \rightarrow W^+ b \rightarrow \ell^+ \nu_\ell\,b$ and the other one decays hadronically $\bar{t} \rightarrow \bar{b} j j$ (also the other way around). The signal contains four jets, one lepton, and missing transverse energy. According to the latest ATLAS $t\bar t +\met$ search~\cite{Aad:2011wc}, the largest SM background after the $\met$ and $M_T$ cuts is $t \bar t$ in the dileptonic channel with one lost lepton and two additional jets from ISR that fake the hadronic $W$. In this section we try to identify some kinematic variables, based on these \emph{background} event topologies.

Before we discuss the new kinematic variables, we first examine the distributions of the signal and main backgrounds in some traditional kinematic variables. In addition to the total missing transverse energy $\met$ and the transverse mass $M_T$\footnote{The transverse mass is defined by the formula $M_T=\sqrt{2p_T^\ell \met [ 1-\cos(\phi^\ell - \phi^{\met}) ] }$, where $p_T^\ell$ is the $p_T$ of the leptons and $\phi^\ell$ and  $\phi^{\met}$ are the azimuthal angles of the lepton and $\vecmet$.} used in the ATLAS analysis~\cite{Aad:2011wc}, we also include the often-used effective mass $m_{\rm eff}$ which is defined as the scalar sum of the four leading jet $p_T$'s, the lepton $p_T$ and $\met$.
Signal and background events are generated using \texttt{MadGraph5}~\cite{Alwall:2011uj}, and showered in
\texttt{PYTHIA}~\cite{Sjostrand:2006za}.  We use \texttt{PGS}~\cite{PGS} to perform the fast detector simulation, after modifying the code to implement the anti-$k_t$ jet-finding algorithm with the distance parameter $R=0.4$~\cite{Cacciari:2008gp}. We simulated the events at 7\,TeV center of mass energy so that we can cross check our results with the ATLAS paper~\cite{Aad:2011wc}.~\footnote{We simulated 6 million events for  the dileptonic and semileptonic $t\bar t$ backgrounds each. Although we only used the unmatched samples in this paper, we checked that a  parton shower plus matrix element matched sample provided good agreement for the basic variable distributions with sufficiently large cuts.} The signal production cross section is normalized to be the value calculated at NLO+NLL~\cite{Beenakker:2011fu}, and the background production cross section for $t\bar t$ is normalized to be the value $\sigma_{t\bar t}(m=173~\mbox{GeV}, 7~\mbox{TeV}) = 163^{+7+9}_{-5-9}$~pb, calculated approximately at NNLO~\cite{Kidonakis:2010dk}. In our studies, the leptonic decays of the top quarks contain $\tau^\pm$ leptons. We adopt the same basic selection cuts on the objects in the final state as in Ref.~\cite{Aad:2011wc} by requiring exactly one isolated electron or muon.

\begin{figure}[h!t]
\begin{center}
\includegraphics[width=0.32\textwidth]{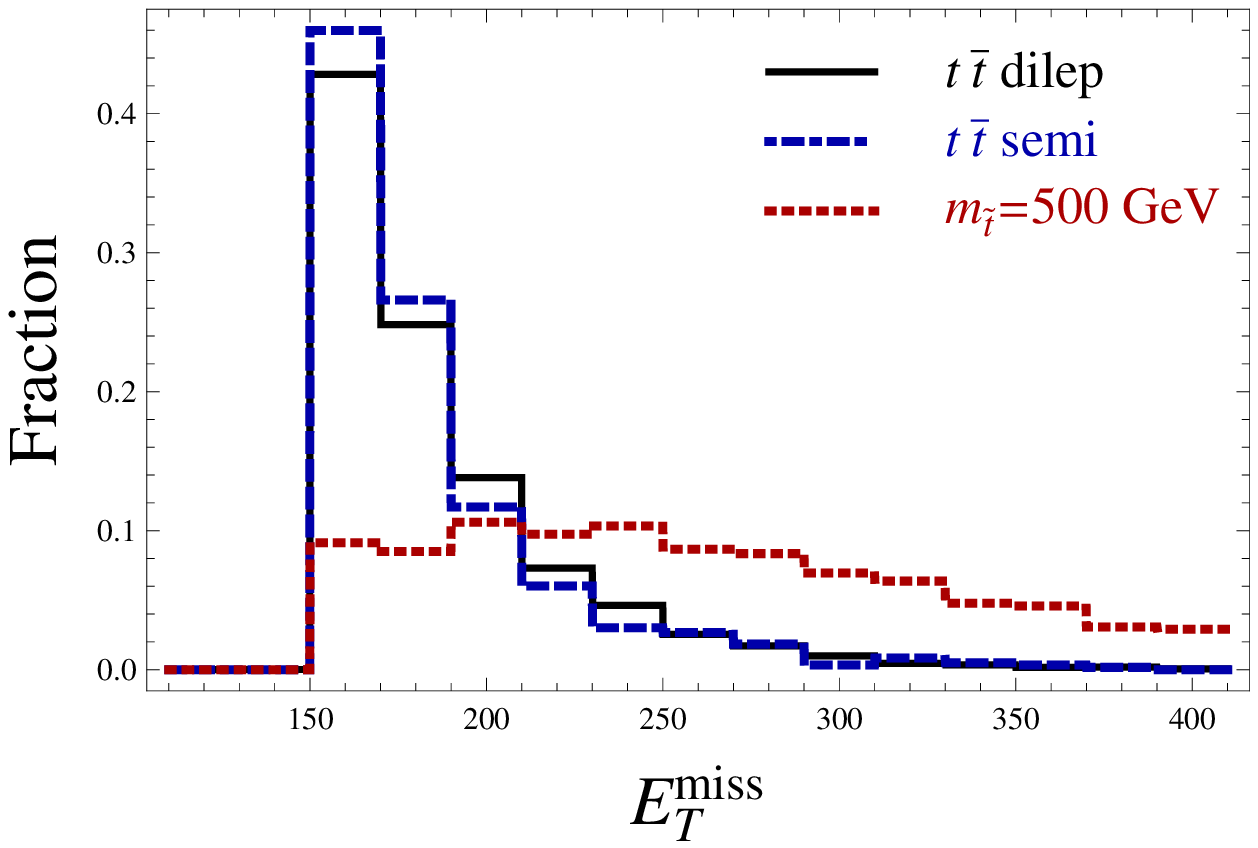}
\includegraphics[width=0.32\textwidth]{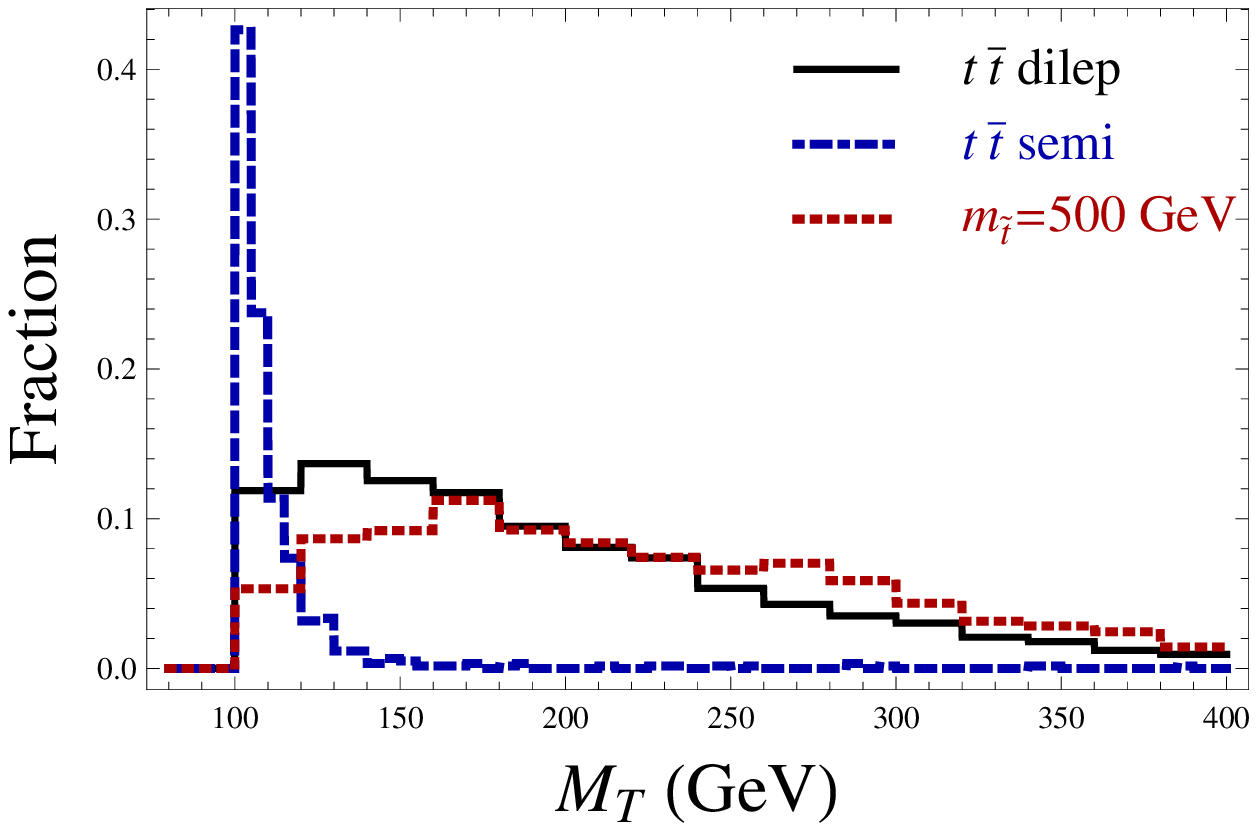}
\includegraphics[width=0.32\textwidth]{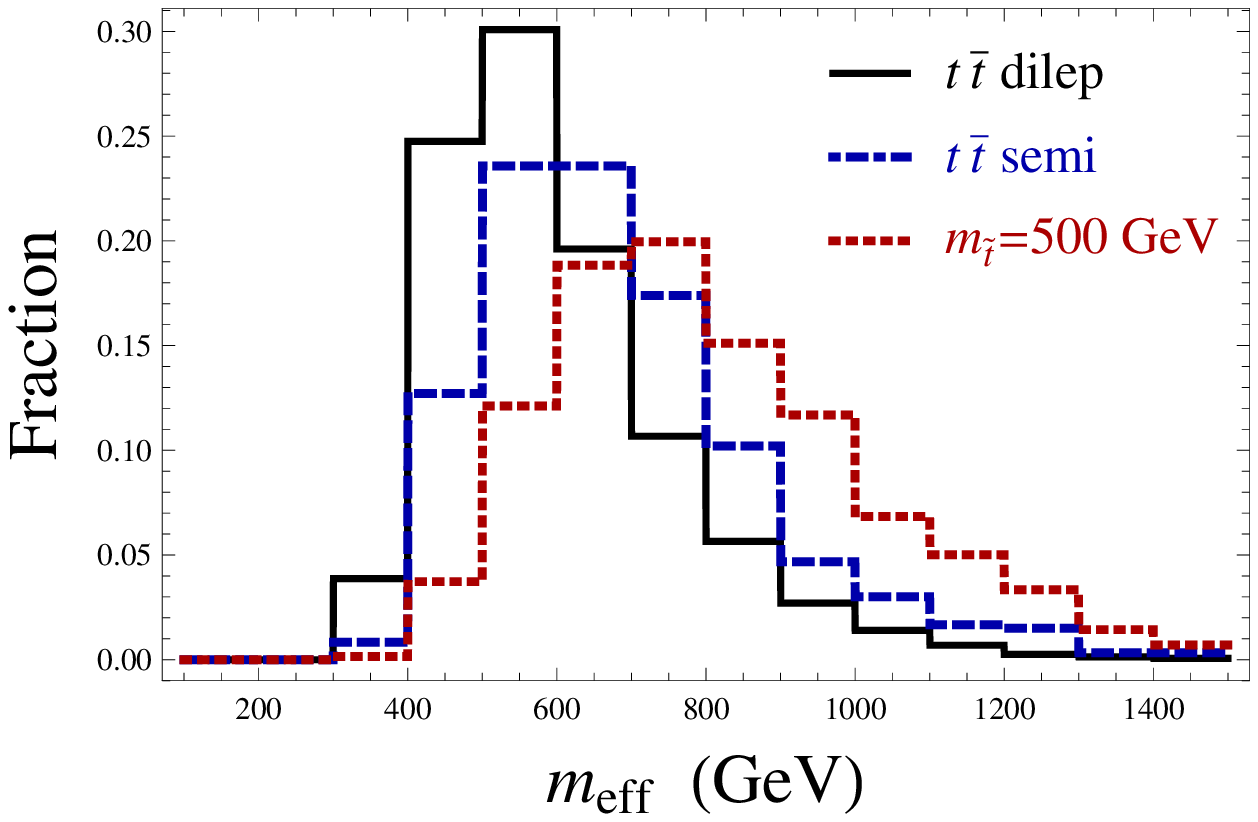}
\caption{The signal and background event distributions in three basic variables: $\met$, $M_T$, and $m_{\rm eff}$. The signal is 7\,TeV production of a 500\,GeV stop pair, each decaying to a top quark and a 100\,GeV neutralino. All the events in the plots have $\met>150$~GeV and $M_T > 100$~GeV.
}
\label{fig:basic}
\end{center}
\end{figure}

Fig.~\ref{fig:basic} shows the signal and background distributions in these three basic variables. For the signal we choose a stop mass of 500\,GeV and the neutralino mass of 100\,GeV. We have included both dileptonic and semileptonic $t\bar{t}$ backgrounds. As one can see from the $M_T$ distributions, the semi-leptonic $t\bar{t}$ background events mainly populate in the region with $M_T < 150$~GeV. Imposing a cut with $M_T > 150$~GeV will be an efficient way to suppress this background. We have also simulated the $W+$jets background and found a similar distribution as the semi-leptonic $t\bar{t}$ background. With the $M_T > 150$~GeV cut, there is only a negligible number of the $W+$jets background events left, so we will not include this background in what follows. The $M_T$ cut is not effective at separating the signal and the dileptonic $t\bar{t}$ background events. On the other hand, cuts on $\met$ and $m_{\rm eff}$ can be used to significantly reduce this background, though it remains the biggest contamination in the direct stop production search.

\begin{figure}[h!t]
\begin{center}
\includegraphics[height=0.25\textheight]{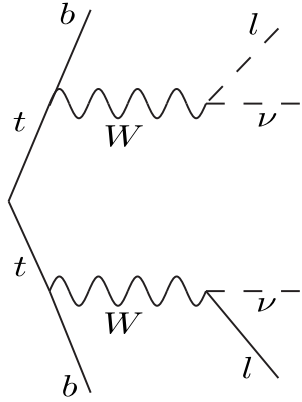}
\caption{The Feynman diagram for the $t \bar t$ background in the (dominant) dileptontic channel. The dashed lines represent missing particles at colliders, including a lost lepton that would otherwise exclude it as a background to our semileptonic stop signal.
}
\label{fig:feyn}
\end{center}
\end{figure}

The diagram for the dileptonic $t\bar{t}$ background event topology is shown in Fig.~\ref{fig:feyn}, with dashed lines representing missing particles. Large $\met$ can arise due to the two missing neutrinos and the missing lepton. Also, the transverse mass $M_T$ is not constrained by the $W$ boson mass because of the additional missing particles. Because there are missing energies on both decay chains, the stransverse mass $M_{T2}$~\cite{Lester:1999tx,Barr:2003rg} can be a natural variable to identify this type of background event.
($M_{T2}$ has been proposed to reduce $t\bar{t}$ and $W^+ W^-$ backgrounds in the di-lepton search channel~\cite{Cohen:2010wv, Kats:2011qh}.)
The $M_{T2}$ for a given event can be interpreted as the minimal mother particle mass compatible with the postulated event topology and an assumed daughter particle mass~\cite{Cheng:2008hk}. The $M_{T2}$ is bounded from above by the mass of the mother particles in the decay chains if the assumed mass for the daughter particles is equal to (or less than) their true mass. By looking at the diagram in Fig.~\ref{fig:feyn}, we can define $M_{T2}$ and its generalizations or variations with the top quark as the mother particle for our backgrounds. Our observables for the leading leptonic background are the 2 $b$-jets + one lepton + $\met$ subsystem. In fact, the next-to-leading dominant semileptonic $t\bar{t}$ background also contains exactly the same subsystem if one disregards the jets from the $W$ decay, so they may be used to bound this background too. On the other hand, the $\tilde{t}\,\tilde{t}^*$ signal has the additional missing energy source from the missing $\tilde{\chi}$ particles. Consequently the corresponding variables can take larger values.

In all $M_{T2}$-type variables, a minimization is performed over all possible ways of dividing $\vecmet$ between the two decay chains. More explicitly, the minimization is over all possible pairs of 4-momenta, each with an assumed mass, whose vector sum has transverse components that match $\vecmet$.  The difference between variables comes in the assignment of visible and missing momentum to the two decay chains, along with invariant mass or $M_T$ constraints imposed on the hidden 4-momenta.  In the following, we define three $M_{T2}$-type variables with background endpoints roughly at the top mass. These new variables are not expected to be completely independent, so their performances will be evaluated in the next section.

\begin{figure}[t]
\begin{center}
\includegraphics[height=0.28\textheight]{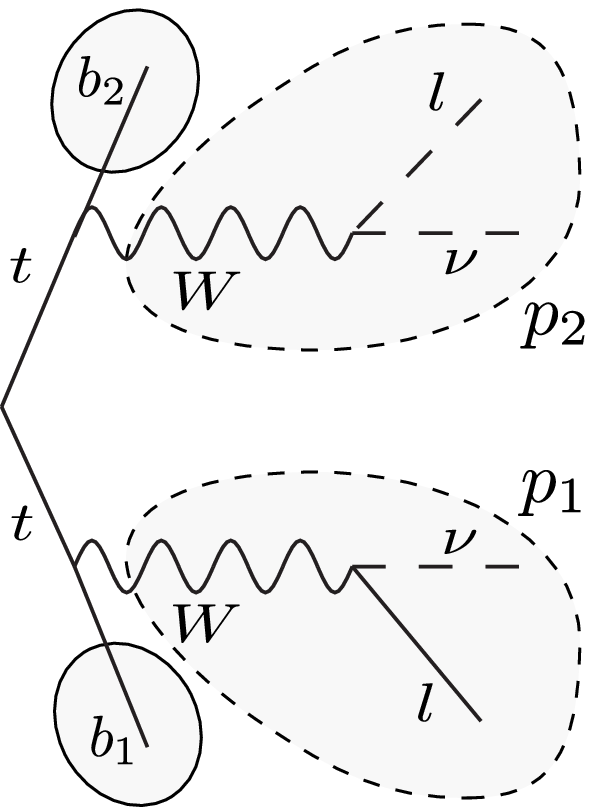}
\hspace{6mm}
\includegraphics[height=0.3\textheight]{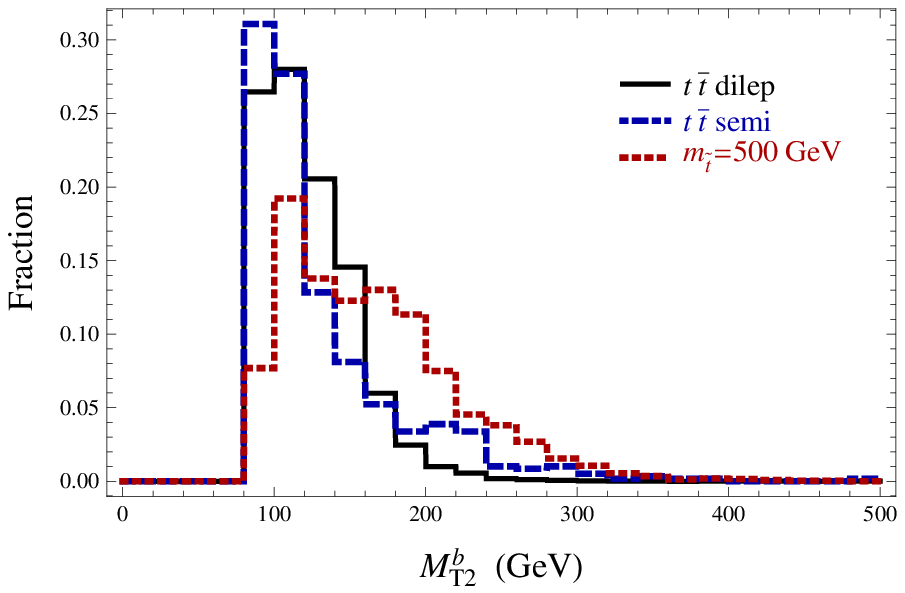}
\caption{
Schematic of $M^{b}_{T2}$, along with its signal and background event distributions. In standard $M_{T2}$ calculations, the $M_T$ of each decay chain is calculated using a visible and invisible momentum. These are indicated by solid and dashed circles in the left panel. The ``invisible on-shell $W$'' momenta, $p_1$ and $p_2$, are scanned in the minimization calculation. The sum of the transverse components is taken to equal $\vecmet$ plus the visible lepton momentum's transverse component. All the events in the plot have $\met>150$~GeV and $M_T > 100$~GeV.}
\label{fig:mt2b}
\end{center}
\end{figure}

The first variable is basically the $M_{T2}$ of the $t\bar{t} \to b W^+ \bar{b} W^-$ subsystem, which is denoted as $M^{b}_{T2}$. Interpreted in the original $M_{T2}$ context, it assumes a ``missing on-shell $W$'' on each side of the decay chain. Since the lepton momentum results from the $W$ decay, we add it to the $\vecmet$. It is defined as
\beqa
M^{b}_{T2} = \mbox{min}\left\{   \bigcup_{ \vec{p}^T_1 + \vec{p}^T_2 = \vecmet + \vec{p}^T_\ell } \mbox{max} {\Big[} M_T(\vec{p}_{b_1},  \vec{p}^T_1), M_T(\vec{p}_{b_2},  \vec{p}^T_2) {\Big]}  \right\}  \, ,
\label{eq:mt2b}
\eeqa
where the $W$ mass is assigned for both $p^T_1$ and $p^T_2$ and jet masses of $p_{b_1}$ and $p_{b_2}$ are calculated from their four-vectors. A diagram illustrating this, along with signal and background distributions of $M^{b}_{T2}$ are shown in Fig.~\ref{fig:mt2b}.
By using the true $W$  boson mass, $M^{b}_{T2}$ is bounded between the top mass and the $W$ gauge boson mass for the dileptonic $t\bar t$ background where this topology is appropriate. (The exact bound holds only for a perfect detector.)

To select the two candidate $b$-jets, we divide all events into three categories. The first category contains exactly two $b$-tagged jets in the four leading jets of $p_T$, and we can use Eq.~(\ref{eq:mt2b}) directly. For the second category containing exactly one $b$-tagged jet, we choose the two leading non-$b$-tagged jets as the other $b$-jet candidate and take the smaller of the two ${M}^b_{T2}$'s. For the third category with zero, three, or four $b$-tagged jets, we assume that the two candidate $b$-jets are contained in the leading three jets and we ignore $b$-tagging information. There are three different combinations, among which we take the smallest as the final value of ${M}^b_{T2}$.

\begin{figure}[h!t]
\begin{center}
\includegraphics[height=0.28\textheight]{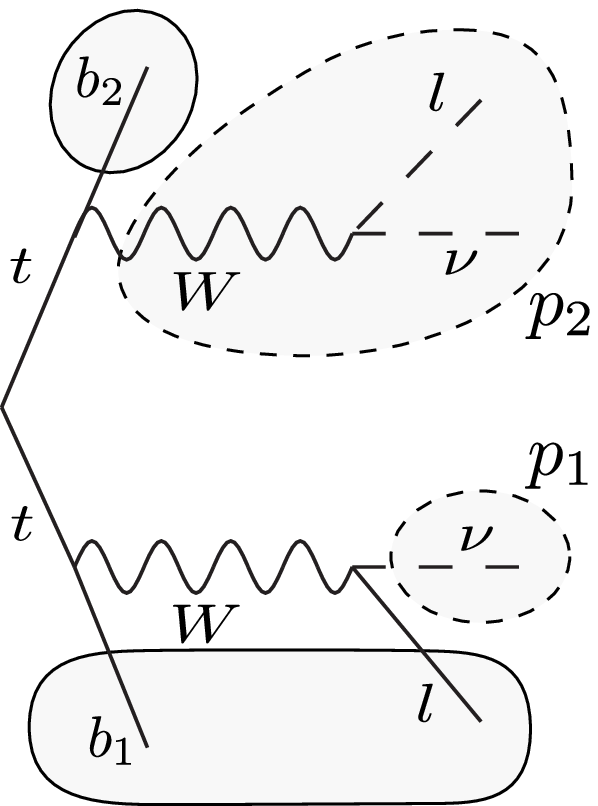}
\hspace{6mm}
\includegraphics[height=0.3\textheight]{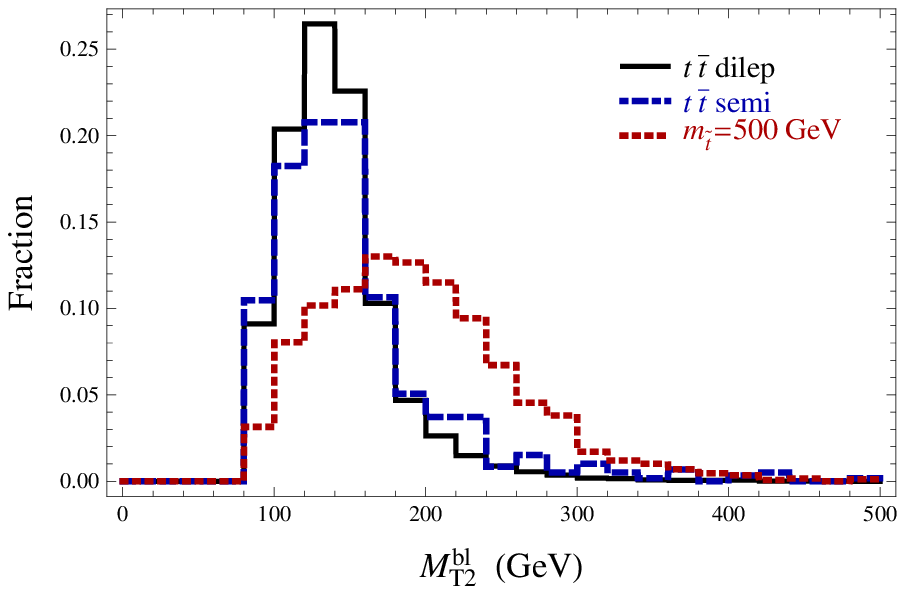}
\caption{ Schematic of $M^{b\ell}_{T2}$, along with its signal and background event distributions. As compared to our previous variable $M^b_{T2}$, the 4-momentum of the visible lepton and $b$-jet are combined together as one effective visible particle, and $p_2$ is treated as an ``invisible on-shell $W$'' when calculating $M_T$ for that side. Again, all  events in the plot have $\met>150$~GeV and $M_T > 100$~GeV.}
\label{fig:mt2bl}
\end{center}
\end{figure}

For the second variable, we do not add the observed lepton momentum to $\met$. Instead we define an asymmetric $M_{T2}$~\cite{Barr:2009jv,Konar:2009qr} by combining the 4-momenta of the lepton and a $b$-jet into one effective particle.  The missing neutrino on that side is treated as massless.  On the other side, the visible particle is the other $b$-jet (with its mass calculated from its four-vector), and the invisible particle is an on-shell $W$.  This variable is defined as
\beqa
M^{b\ell}_{T2} = \mbox{min}\left\{   \bigcup_{\vec{p}^T_1 + \vec{p}^T_2 = \vecmet } \mbox{max} {\Big[} M_T(\vec{p}_{b_1} + \vec{p}_{\ell},  \vec{p}^T_1), M_T(\vec{p}_{b_2},  \vec{p}^T_2) {\Big]} \right\}  \,.
\label{eq:mt2bl}
\eeqa
The two $b$-jet candidates are chosen by the same procedure as in the previous case. There are two ways to pair the lepton with one of the two $b$-jets, and  the combination which produces a smaller $M^{b\ell}_{T2}$ is chosen.
A diagram illustrating the calculation, along with signal and background distributions are shown in Fig.~\ref{fig:mt2bl}.
%


Each of the two $M_{T2}$ variables defined above did not fully utilize the information available for the background event topology: the two intermediate $W$ bosons are on-shell and one of them produces the observed lepton together with a neutrino. We can define a new kinematic variable as the minimal mother particle mass (the top quark mass in this case) which can be compatible with all the transverse momentum and mass-shell constraints of that topology for a given event. Here, the top quark mass is not explicitly used, only implicitly bounded by the event. This is in the same spirit as interpreting $M_{T2}$ as the minimal mother particle mass compatible with the minimal kinematic constraints~\cite{Cheng:2008hk} except that all mass-shell constraints on the cascade decay chain are used~\footnote{The mass-shell constraints are not sufficient to fully reconstruct each event.}. One might expect to get a variable which is more sensitive to this background topology because of the additional kinematic information applied in the definition. Specifically, the variable $M^W_{T2}$ (where the superscript $W$ represents the on-shell intermediate $W$ information is included when combining lepton and neutrino) can no longer be cast into the ``maximum of two side's $M_T$'' form, but is instead defined directly as the minimization~\footnote{The programs for calculating all new variables defined in this paper can be downloaded at \url{https://sites.google.com/a/ucdavis.edu/mass/}}
\beqa
M^W_{T2} &=&  \mbox{min}
\left\{
  m_{y} \mbox{ consistent with: } \left[ \begin{array}{r}
\vec{p}^T_1 + \vec{p}^T_2 =  \vecmet\,,\;  p_1^2=0  \,,\;
(p_1 + p_{\ell} )^2 = p_2^2 = M_W^2 \,,\;\\
(p_1 + p_{\ell} + p_{b_1})^2=(p_2 + p_{b_2})^2=m^2_y \,\;
\end{array}\right]
\right\} .
\label{eq:MT2W}
\eeqa

\begin{figure}[h!t]
\begin{center}
\includegraphics[height=0.28\textheight]{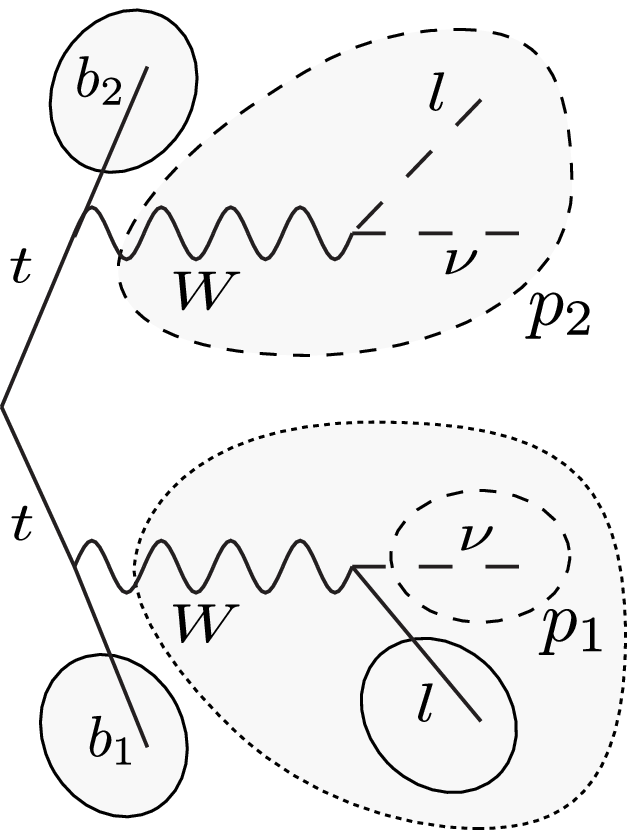}
\hspace{6mm}
\includegraphics[height=0.3\textheight]{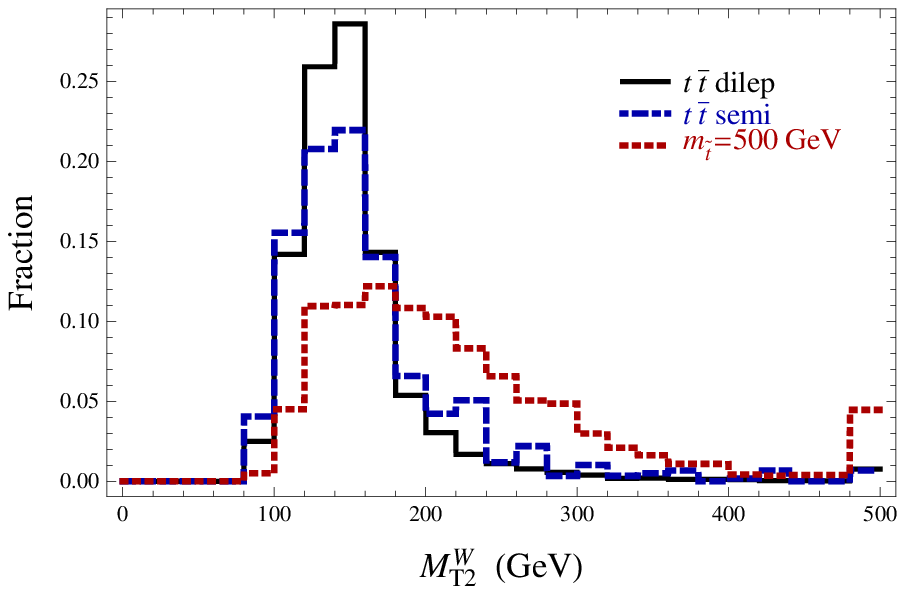}
\caption{Schematic of $M^W_{T2}$, along with its signal and background event distributions.   Here all of the information is used, including the $W$-on-shell mass condition on both sides.  As with the other variables, $p_2$ is the entire missing on-shell $W$, but $p_1$ is the neutrino that gets paired with the visible lepton to form the other on-shell $W$.  All the events in the plot have $\met>150$~GeV and $M_T > 100$~GeV. The events with no compatible top mass under 500\,GeV are placed in the last bin.}
\label{fig:mt2w}
\end{center}
\end{figure}

The diagram, along with signal and background distributions are shown in Fig.~\ref{fig:mt2w}. We use the same method as before to pick the two $b$-jets, and a method similar to that for $M^{b\ell}_{T2}$ is used to choose which $b$-jet gets paired with the visible lepton.
Calculating this variable can be done efficiently in a similar way as the $M_{T2}$ calculation in Ref.~\cite{Cheng:2008hk} by generalizing the method there to this case. For perfect measurements, this variable for the dileptonic $t\bar{t}$ backgrounds is less than the true top quark mass since the top mass should be compatible with all background events.  On the other hand, the signal events do not need to satisfy such a bound, because of its different topology and additional missing massive particles $\tilde{\chi}$.  For some of the signal events we may not even be able to find a compatible mass because we apply the variable to a wrong topology with the wrong mass-shell conditions. The background distributions indeed lie mostly below the top quark mass, while a significant number of signal events have no solution below 500~GeV and they are included in the last bin.

One can see from the plots in Figs.~\ref{fig:mt2b}, \ref{fig:mt2bl}, and \ref{fig:mt2w} that a cut on these variables around the top quark mass could be an effective way to suppress the main background. It is not clear {\it a priori} which one will have the best performance when the experimental smearing and detector resolution effects are taken into account, and whether there is still enough independent information among them so that a combination of them can give some further improvement. In the next section we will make a critical comparison of the performances of these variables and their combinations.

\section{Performances of New Kinematic Variables}
\label{sec:performance}

To quantify the power of these kinematic variables, we optimized a simple cut-and-count experiment involving three stop masses (400, 500, 600) GeV, with the neutralino mass being fixed at 100\,GeV and 100\% branching ratio of stop decaying to top plus neutralino. We simulated the signal and background events at 7\,TeV to compare with the existing ATLAS study. Although the LHC will run at 8\,TeV this year, the relative performance of each kinematic variable will not be affected much. We will comment on the 8\,TeV case in the next section.
For each possible set of cuts, we use the NLO cross sections of stop pair productions multiplied by cut efficiencies to estimate the number of signal $(S)$ and background $(B)$ events expected in 20\,fb$^{-1}$ of data.  The Poisson probability that pure background would fluctuate up to at least $S+B$ events is given by {\tt ROOT}'s two-parameter $\Gamma$ function, which can be evaluated at non-integer parameters:
\beqa
p \ = \  \sum_{k=S+B}^\infty \frac{B^k}{k!} e^{-B} \ =\  {\tt TMath::Gamma}(S+B, B) .
\eeqa
We translate this probability into a gaussian-equivalent significance ($\sigma$) in terms of standard deviations.\footnote{This translation is also handled by {\tt ROOT} as $\sigma \ = \ {\tt TMath::NormQuantile}( 1 - p )$.}  This approaches $S/\sqrt{B}$ for large signal and background, but by handling the small-number statistics, we avoid extreme cuts.
By finding cuts that maximize this significance reach, we can estimate the power of of any set of kinematic variables.

To evaluate the performances of the new kinematic variables defined in the previous section, we include them with a basic set of cuts on $(\met,\, M_T, \, m_{\rm eff})$. Within the basic set of variables, the $\met$ is most powerful in discriminating the signal and the background. A cut on $M_T>150$\,GeV is imposed to remove the $W+$jets background, and after this cut, the semileptonic $t\bar{t}$ background is virtually eliminated. We found that further increasing the $M_T$ cut will hurt the signal significance, so we fixed the $M_T$ cut to be at 150\,GeV in our study. The effective mass $m_{\rm eff}$ is not as useful as $\met$, and its inclusion will only slightly improve the results. However, it is a simple variable and is widely used, so we still include it in the optimization. We added the new kinematic variables ($M^b_{T2},\, M^{b\ell}_{T2},\, M^W_{T2}$) one at a time to the basic set  and compare improvements by re-optimizing the cuts on all variables. We also combined all three new variables with the basic set to see if there is any independent information among these new variables which can further improve the significance. The results are listed in Tables~\ref{table:stop400} and \ref{table:stop500}.

\begin{table}
\begin{center}
\begin{tabular}{|c|c|c|c|c||r|r|c|c|}
\multicolumn{5}{c}{Minimum Cuts}  &
\multicolumn{4}{c}{$m_{\rm stop}=400$\,GeV}   \\
\hline
$\met$   &    $m_{\rm eff}$   &    $M^W_{T2}$   &    $M_{T2}^b$   &    $M_{T2}^{bl}$   &    $S_{20fb^{-1}}$  &  $B_{20fb^{-1}}$   &  $S/B$  &   $\sigma$  \\
\hline
(150)  &   -  &   -  &   -  &   -  &   129.3  &  738.4  &   0.17  &   4.62  \\
202  &   -  &   -  &   -  &   -  &   82.1  &  208.8  &   0.39  &   5.34  \\
202  &   491  &   -  &   -  &   -  &   81.7  &  202.7  &   0.40  &   5.38  \\
202  &   502  &   -  &   100  &   -  &   74.4  &  147.9  &   0.50  &   5.66  \\
202  &   502  &   -  &   -  &   157  &   55.9  &  66.7  &   0.84  &   6.09  \\
200  &   562  &   177  &   -  &   -  &   50.7  &  48.5  &   1.05  &   6.33  \\
200  &   564  &   176  &   95  &   99  &   50.0  &  46.1  &   1.08  &   6.38  \\
\hline
\end{tabular}
\end{center}
\caption{Cuts optimized for significance to discover 400\,GeV stop and 100\,GeV neutralino for 20\,fb$^{-1}$ at 7\,TeV. Runs optimized over different cut variables are sorted by increasing significance (roughly $S/\sqrt{B}$ for large numbers of events, where the improvement factor becomes independent of integrated luminosity.)  All optimizations began with $\met>150$\,GeV (first row) and include a fixed $M_T>150$\,GeV cut to eliminate $W+$jets. The starting cuts yield 12120 simulated $t\bar t$ dileptonic and 24 semileptonic background events, and 972 simulated signal events. The numbers of the signal and background events in the table are rescaled to 20 fb$^{-1}$ luminosity.  The $M_T$ cut was for the $W$+jets background, and increasing it beyond  150\,GeV never helped with the $t \bar t$ backgrounds, so its column is not shown.}
\protect\label{table:stop400}
\end{table}

From these results we see that indeed the new kinematic variables can improve the signal significance on top of the basic variables. (The effective mass $m_{\rm eff}$, on the other hand, does not help very much.) For the $400$~GeV stop, $M^W_{T2}$ has the best performance and $M^b_{T2}$ is the least useful one among the three variables. This is in accordance with our expectation of these three, as the variable $M^W_{T2}$ contains the most kinematic information of the background event topology while $M^b_{T2}$ contains the least. For a heavier stop of 500 or 600 GeV, the performances of the three variables are actually comparable. (One should not take the small differences seriously due to the limited statistics.) These new variables are highly correlated and not much improvement can be gained by combining all of them.  Other than improvement on the discovery sensitivity via $S/\sqrt{B}$, we also note that from Tables~\ref{table:stop400} and \ref{table:stop500} the improvement on $S/B$ is even more dramatic after including the new variables defined in this paper. So, the systematic errors for the actual experimental searches can further reduced.

\begin{table}
\begin{center}
\begin{tabular}{|c|c|c|c|c||r|r|c|c|}
\multicolumn{5}{c}{Minimum Cuts}  &
\multicolumn{4}{c}{$m_{\rm stop}=500$\,GeV}   \\
\hline
$\met$   &    $m_{\rm eff}$   &    $M^W_{T2}$   &    $M_{T2}^b$   &    $M_{T2}^{bl}$   &    $S_{20fb^{-1}}$  &  $B_{20fb^{-1}}$   &  $S/B$  &   $\sigma$  \\
\hline
(150)  &   -  &   -  &   -  &   -  &   34.0  &  738.4  &   0.05  &   1.23  \\
303  &   -  &   -  &   -  &   -  &   11.4  &  16.6  &   0.69  &   2.49  \\
303  &   659  &   -  &   -  &   -  &   11.4  &  16.1  &   0.70  &   2.50  \\
299  &   709  &   172  &   -  &   -  &    9.8  &   6.2  &   1.59  &   3.19  \\
291  &   743  &   -  &   163  &   -  &    7.9  &   3.6  &   2.21  &   3.20  \\
300  &   708  &   -  &   -  &   170  &    9.4  &   5.6  &   1.69  &   3.20  \\
291  &   742  &   173  &   123  &   109  &    9.0  &   4.4  &   2.04  &   3.34  \\
\hline
\end{tabular}

\vspace{2em}

\begin{tabular}{|c|c|c|c|c||r|r|c|c|}
\multicolumn{5}{c}{Minimum Cuts}  &
\multicolumn{4}{c}{$m_{\rm stop}=600$\,GeV}   \\
\hline
$\met$   &    $m_{\rm eff}$   &    $M^W_{T2}$   &    $M_{T2}^b$   &    $M_{T2}^{bl}$   &    $S_{20fb^{-1}}$  &  $B_{20fb^{-1}}$   &  $S/B$  &   $\sigma$  \\
\hline
(150)  &   -  &   -  &   -  &   -  &   16.7  &  738.4  &   0.02  &   0.60  \\
377  &   -  &   -  &   -  &   -  &    4.5  &   3.0  &   1.49  &   2.04  \\
345  &   696  &   -  &   -  &   -  &    6.1  &   6.3  &   0.97  &   2.05  \\
337  &   727  &   168  &   -  &   -  &    5.9  &   3.0  &   2.01  &   2.66  \\
337  &   726  &   -  &   -  &   168  &    5.8  &   2.7  &   2.17  &   2.69  \\
333  &   740  &   -  &   157  &   -  &    5.3  &   2.1  &   2.59  &   2.73  \\
332  &   741  &   168  &   148  &   91  &    5.5  &   2.1  &   2.67  &   2.81  \\
\hline
\end{tabular}
\end{center}
\caption{Cuts optimized for significance to discover 500\,GeV and 600\,GeV stops with 100\,GeV neutralinos for 20\,fb$^{-1}$ at 7\,TeV. Again, all runs began with $\met>150$\,GeV and include a fixed $M_T>150$\,GeV cut (not shown), where there are 2115 and 1938 simulated events for 500 GeV and 600 GeV stops and the same number of background events as in Table~\ref{table:stop400}.  Cuts on $\met$ and $M^W_{T2}$ still do almost as well as optimization over all variables, but here these additional cuts can improve $S/B$.}
\protect\label{table:stop500}
\end{table}

We also tried a few small variations of these new variables and did not obtain better results.
For example, in $M^b_{T2}$, using zero mass for $W$, or not adding the lepton momentum to $\met$ yields very similar results, and these variations are more than 95\% correlated. Assuming the transverse momentum of the missing neutrino from the $W$ decay is in the same direction as that of the observed lepton gives a worse result, since the $W$ bosons in the background events are in general not highly boosted. An $M_{T2}$ variable motivated by the signal topology by combining one $b$-jet with the lepton and the other $b$-jet with two non-$b$-jets also does not help. Our results indicate that if one wants to choose a minimal set of variables for the semileptonic channel search of the stop direct production for a wide range of the stop mass, the set $(\met, \, M_T,\, M^W_{T2})$ (even without $m_{\rm eff}$) can achieve nearly the maximal discriminant power of combining many different variables.

\section{Conclusions}
\label{conclusions}

The LHC will be running at 8\,TeV in 2012 and is anticipated to achieve a larger integrated luminosity. We expect an even higher mass reach for the stop search compared with the numbers obtained in the previous section based on 7\,TeV. To estimate the exclusion or discovery sensitivity of the stop with a 20\,fb$^{-1}$ luminosity, we calculated the stop signal cross sections at tree level using  \texttt{MadGraph5} and applying the same $K$-factor at the 7\,TeV LHC to take into account the QCD NLO corrections. The same procedure is applied to the $t \bar t$ backgrounds to obtain the approximate NLO production cross section at the 8\,TeV LHC. The total $t \bar t$ cross section is calculated to be 231.8\,pb. With the help of the new kinematic variables discussed in this paper together with the basic variables $\met$, $M_T$ and $m_{\rm eff}$, we found that for $m_{\tilde t} = 650$~GeV and $m_{\chi^0}=100$~GeV with $\mbox{Br}(\tilde t \rightarrow t + \chi^0) =100\%$, the stop can show up at the $4\sigma$ level if we ignore the systematic errors. The 95\% C.L. exclusion reach can go up to around 700~GeV. If there is no excess found in the 8\,TeV run, it will dent the hope of a non-fine-tuned SUSY solution to the hierarchy problem, unless the stop has some more exotic signatures in some non-standard scenarios, such as degenerate spectrum or $R$-parity violation, etc.

In this paper we have focused on suppressing the $t\bar{t}$ backgrounds for the search of direct stop production. However, the Standard Model $t\bar{t}$ production is a major background for a wide range of new physics searches at the LHC. The kinematic variables proposed here could also be useful in improving searches for other new physics where the $t\bar{t}$ constitutes the main background with a large missing transverse momentum for the signals.

Comparing the performances of different variables shows that in general the more kinematic information a variable contains, the more discriminant power it can possess. For more specific new physics searches where both the signal and main background event topologies are known, it is worth designing kinematic variables which carry as much information of the signal and/or the background events as possible to achieve the maximal discrimination between them. The strategy of constructing new kinematic variables discussed in this paper could be readily generalized to other cases.

\subsection*{Acknowledgments}
We would like to thank the collaboration of Markus Luty in the early stage of this work, and Michael Peskin for useful discussion and comments. This work was supported in part by U.S. DOE grant No. DE-FG02-91ER40674. SLAC is operated by Stanford University for the US Department of Energy under contract DE-AC02-76SF00515.

 \end{document}